\documentclass[twocolumn]{aastex631}

\usepackage{amsmath}
\received{December 30, 2021}
\accepted{March 25, 2022}

\shorttitle{Extended millimeter emission in 3C273}
\shortauthors{Komugi et al.}
\graphicspath{{./}{figures/}}

\begin{document}

\title{Detection of extended millimeter emission in the host galaxy of 3C273 and its implications for QSO feedback via high dynamic range ALMA imaging}

\correspondingauthor{Shinya Komugi}
\email{skomugi@cc.kogakuin.ac.jp}

\author{Shinya Komugi}
\affiliation{Division of Liberal Arts, Kogakuin University, 2665-1 Nakano-cho, Hachioji, Tokyo 192-0015, Japan}

\author{Yoshiki Toba}
\affiliation{Department of Astronomy, Kyoto University, Kitashirakawa-Oiwake-cho, Sakyo-ku, Kyoto 606-8502, Japan}
\affiliation{Academia Sinica Institute of Astronomy and Astrophysics, 11F of Astronomy-Mathematics Building, AS/NTU, No.1, Section 4, Roosevelt Road, Taipei 10617, Taiwan}
\affiliation{Research Center for Space and Cosmic Evolution, Ehime University, 2-5 Bunkyo-cho, Matsuyama, Ehime 790-8577, Japan}

\author{Yoshiki Matsuoka}
\affiliation{Research Center for Space and Cosmic Evolution, Ehime University, 2-5 Bunkyo-cho, Matsuyama, Ehime 790-8577, Japan}

\author{Toshiki Saito}
\affiliation{Department of Physics, General Studies, College of Engineering, Nihon University, 1 Nakagawara, Tokusada, Tamuramachi, Koriyama, Fukushima 963-8642, Japan}
\affiliation{National Astronomical Observatory of Japan, 2-21-1 Osawa, Mitaka, Tokyo 181-8588, Japan}

\author{Takuji Yamashita}
\affiliation{Research Center for Space and Cosmic Evolution, Ehime University, 2-5 Bunkyo-cho, Matsuyama, Ehime 790-8577, Japan}
\affiliation{National Astronomical Observatory of Japan, 2-21-1 Osawa, Mitaka, Tokyo 181-8588, Japan}



\begin{abstract}
We estimate the amount of negative feedback energy injected into the ISM of the host galaxy of 3C273, a prototypical radio loud quasar.  We obtained 93, 233 and 343\ GHz continuum images with the Atacama Large Millimeter/Sub-millimeter Array (ALMA).  After self calibration and point source subtraction, we reach an image dynamic range of $\sim 85000$ at 93\ GHz, $\sim 39000$ at 233\ GHz and $\sim 2500$ at 343\ GHz.  These are currently the highest image dynamic range obtained using ALMA.  We detect spatially extended millimeter emission associated with the host galaxy, cospatial with the Extended Emission Line Region (EELR) observed in the optical.  The millimeter spectral energy distribution and comparison with centimeter data show that the extended emission cannot be explained by dust thermal emission, synchrotron or thermal bremsstrahlung arising from massive star formation.  We interpret the extended millimeter emission as thermal bremsstrahlung from gas directly ionized by the central source.  The extended flux indicates that at least $\sim 7\%$ of the bolometric flux of the nuclear source was used to ionize atomic hydrogen in the host galaxy.  The ionized gas is estimated to be as massive as $10^{10}$ to $10^{11}\ \mathrm{M_\odot}$, but the molecular gas fraction with respect to the stellar mass is consistent with other ellipticals, suggesting that direct ionization ISM by the QSO may not be sufficient to suppress star formation, or we are witnessing a short timescale before negative feedback becomes observable. The discovery of a radio counterpart to EELRs provides a new pathway to studying the QSO-host ISM interaction.
\end{abstract}



\section{Introduction} \label{sec:intro}

The question of how much of the energy from a QSO is fed back into its host galaxy remains to be one of the most unanswered questions regarding galaxy evolution.  While negative feedback is necessary to provide star formation quenching and thus explain the current number of star forming galaxies in the nearby universe, current observational evidence of negative feedback give varied results.  Cold molecular outflows are found in a number of galaxies \citep{garcia15, salak16, veilleux17, salak20, garcia21} which could contribute to quenching by depriving the galaxy of fuel for star formation.  The effect of the outflow may be relatively limited to nearby the AGN, however \citep{garcia14}, and in special cases, star formation can occur in outflows as well \citep{maiolino17}.

Outflows may also come in the form of hot gas \citep{cano-diaz12, zakamska16b, toba17a}, which may \citep{garcia14} or may not \citep{toba17b} be coupled with cold molecular gas from the galaxy.  Simulations show that hot outflows driven by AGN can prevent cooling flows and work as a preventive feedback \citep{biernacki18}, but special cases have been found where hot gas may eventually form molecular gas and contribute to star formation in the disk \citep{russell17}.  It is not clear whether outflows are ubiquitous, and if they are responsible for stopping star formation in the bulk of galaxies.

\begin{figure*}[htb]
\includegraphics{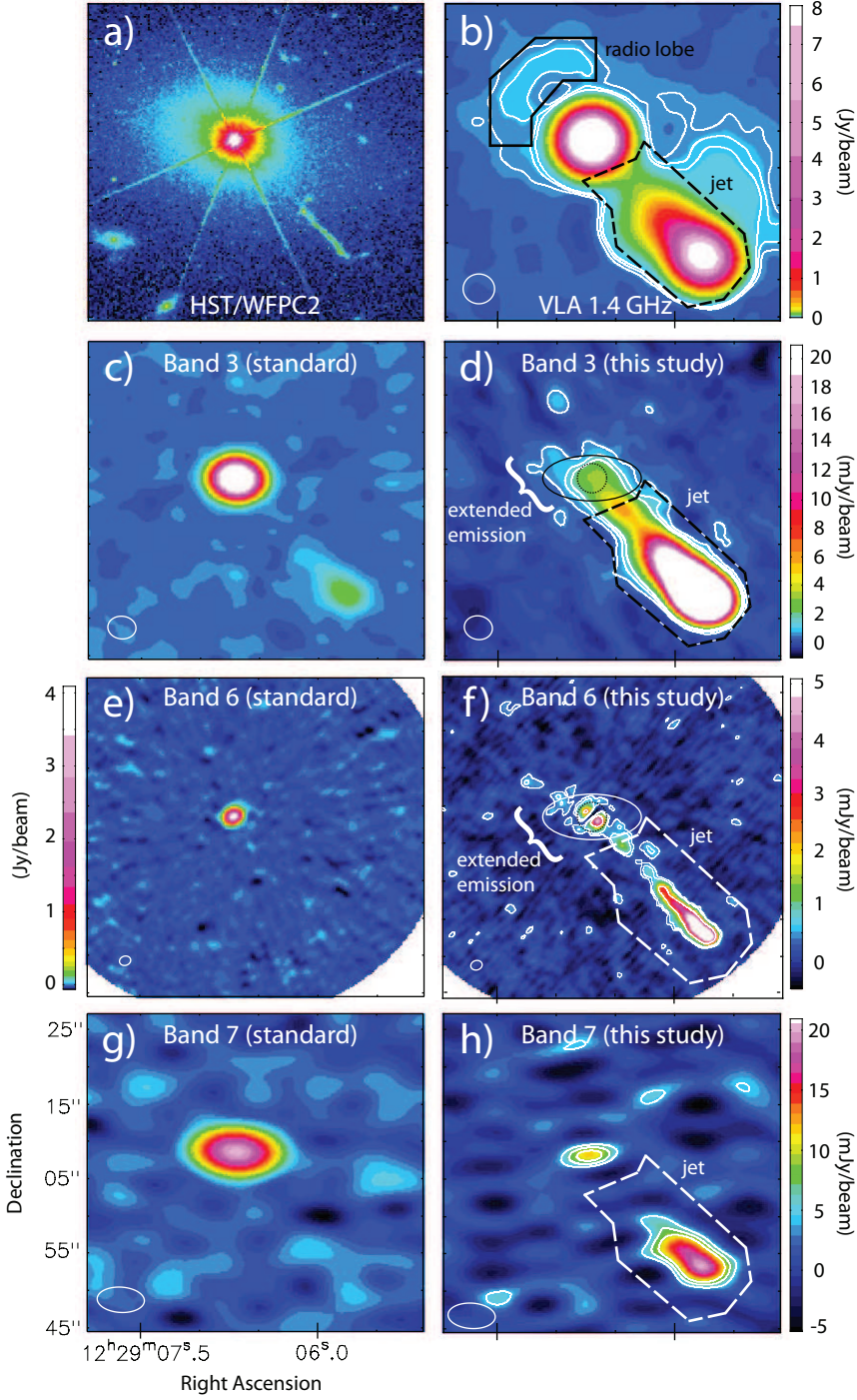}
\caption{a) HST/WFPC2 F606W image$^1$ of 3C273.  The host elliptical is seen as a diffuse structure lopsided to the northeast.  The jet structure extends to the southwest.  b) VLA 1.4 GHz image from \citet{perley17}.  Contours are drawn at $1.0$, $1.5$ and $2.0\times 7\mathrm{mJy\ beam^{-1}}$, where $\mathrm{beam}=4^{\prime \prime}.5$.  Colorbar is shown on the right side.  Also shown in solid black lines are the regions inside which the radio lobe flux estimation (section \ref{synch1}) and jet 
flux measurement (section \ref{jet}) were done.  c) ALMA Band 3 (93 GHz) image of 3C273 using the standard reduction procedure.  Common colorbars are
used for panels c), e) and g), drawn left of panel e).  The Band 3 beam is shown in the bottom left corner.  d) Band 3 after self calibration and point source subtraction explained in section \ref{selfcal} and \ref{pssub}.  Solid contour is drawn at $5$, $10$ and $15\sigma$, where $1\sigma = 0.11\ \mathrm{mJy\ beam^{-1}}$.  The solid ellipse is the $13^{\prime \prime}\times 6^{\prime \prime}$ ellipse where our extended emission flux were measured.  Dashed line shows the $4^{\prime \prime}$ region where flux inside was subtracted from measurements, due to reasons explained in section \ref{section_b6}.
e)  Same as c) for ALMA Band 6 (230GHz).  The Band 6 beam is shown in the bottom left corner.  f)  Same as d) for Band 6.  Solid contours is drawn at $3$, $5$ and $20\sigma$, where $1\sigma = 0.12\ \mathrm{mJy\ beam^{-1}}$.  g) Same as c) for ALMA Band 7 (340GHz).  The Band 7 beam is shown in the bottom left corner.  h)  Same as d) for Band 7.  Contours in the are drawn at $3$,$5$ and $7\sigma$, where $1\sigma = 1.3\ \mathrm{mJy\ beam^{-1}}$.
\label{image}}
\end{figure*}

Another scenario for stopping star formation in the host galaxy is direct ionization of gas in the host galaxy by the nuclear source.  Extended Emission Line Regions (EELRs) have been discovered in galaxies hosting QSOs, 
some of which can extend to tens of kiloparsecs \citep{stockton02, yoshida02, husemann08, fu09}.  It has been difficult to quantitatively discuss how much energy from the QSO is fed back to the host, however, due to the numerous assumptions necessary to interpret the EELRs.  The EELRs are commonly detected using forbidden optical emission lines like [OIII]$\lambda$5007 \citep{matsuoka12, husemann19a, husemann19b}, whose energetics require assumptions regarding excitation conditions, and are susceptible to dust absorption as well.  
An alternative method of quantifying negative feedback is to detect the thermal bremsstrahlung arising from gas directly ionized by the nuclear source.  Thermal bremsstrahlung is detected at millimeter wavelengths, and thus free from dust extinction.  The simplicity of the excitation mechanism requires less assumptions. It has been difficult to detect extended radio emission in the host galaxy of QSOs, however, as the brightness of the central QSO and its associated sidelobes normally swamps any faint emission.  Only if we obtain exceptionally high image dynamic range (e.g., \citet{punsly16}), we may expect to detect such thermal emission.
In this paper we present millimeter observations of the prototypical radio loud QSO 3C273 (panel a) of figure \ref{image}) at $z=0.158$ \citep{schmidt63} as a testbed for this method.  3C273 is the brightest radio source in the sky, so that if faint millimeter emission can be detected via high dynamic range imaging in this source, the method would be applicable to any QSO, and may become a powerful alternative to quantifying energetics in the interaction between QSO and their host galaxy ISM.

In this paper we assume a $\Lambda$CDM cosmology with $\Omega_m = 0.27$, $\Omega_\Lambda = 0.73$, and $H_0=71\ \mathrm{km\ s^{-1}\ Mpc^{-1}}$ \citep{spergel03}.  These assumptions give a luminosity distance of 749 Mpc to 3C273, and $1^{\prime \prime}$ corresponds to a projected scale of $2.7$ kpc.

\footnote{Based on observations made with the NASA/ESA Hubble Space Telescope, and obtained from the Hubble Legacy Archive, which is a collaboration between the Space Telescope Science Institute (STScI/NASA), the Space Telescope European Coordinating Facility (ST-ECF/ESA) and the Canadian Astronomy Data Centre (CADC/NRC/CSA).
}

\section{Observation and Data Reduction} \label{subsec:tables}

\subsection{Band 3}
3C273 was observed at 93 GHz (ALMA Band 3) as a bandpass and phase calibrator for Cycle 6 (2017.1.00815).  
The observations were conducted over 3 scans on 3-4 June 2018 and  6 scans on 16-19 January 2019.  Each scan lasted 5 minutes.  

The raw data in ASDM (ALMA Science Data Model) format were downloaded from the ALMA archive and re-processed using the standard pipeline (version 5.6.1-8).  After additional flagging at spectral window edges, imaging was performed using $tclean$ with natural weighting.  The resulting peak flux was 9.37 $\mathrm{Jy/beam}$ with a standard deviation of $1\sigma = 6.1\ \mathrm{mJy/beam}$ at emission free regions, and a beam size of $3^{\prime \prime}.6 \times 3^{\prime \prime}.0$.  Panel c) of figure \ref{image} shows this ``standard'' image, clearly indicating that the image noise is dominated by sidelobes of the bright central source, i.e., that the image noise is limited by dynamic range (DR).  Standard ALMA images are expected to reach DR of $\sim 100$ for lower frequency bands (ALMA Cycle 8 Proposers Guide, Doc. 8.2., ver. 1.0), which is clearly insufficient to detect any weak emission associated with the host galaxy.  We attempted to increase the DR via the following steps.

\subsubsection{Self Calibration}\label{selfcal}
3C273 can be modeled with a simple point source, thus can be used as a model to calibrate itself.  QSO flux can change considerably over a timescale of several months.  The peak flux of 3C273 in the 2018 run was measured to be 12.2 $\mathrm{Jy/beam}$ ($1\sigma = 3.20\ \mathrm{mJy/beam}$), decreasing to 7-8 $\mathrm{Jy/beam}$ ($1\sigma = 0.77\ \mathrm{mJy/beam}$) in the 2019 run.
Therefore, we performed self-calibration separately for the 2018 and 2019 runs.  For both runs, the solution interval was set to 1 second and solved for both phase and amplitude.  For the 2018 run, the image noise changed only slightly, to $1\sigma = 3.16\ \mathrm{mJy/beam}$, but for the 2019 run the noise decreased to $1\sigma = 0.41\ \mathrm{mJy/beam}$.  Changes in the peak flux were negligible for both runs.

\subsubsection{Point Source Subtraction} \label{pssub}
Modeling and subtracting the central point source, which is responsible for the large sidelobes, is important to increase the image DR.  This was done using the following steps.
First, the weights of the visibility of the 2018 and 2019 runs were normalized using the CASA task $statwt$.  We performed $tclean$ separately for each scans, with the $savemodel = ``modelcolumn" $ parameter and the clean box set to only the central peak pixel in order to save the central component as a model.  The channel width was chosen to be $10\ \mathrm{MHz}$, as a reasonable compromise between averaging out phase variations and preserving the small scale bandpass characteristics of the point source.  The resulting point source model has a frequency resolution of $10\ \mathrm{MHz}$ and time resolution of 5 minutes (i.e., the length of 1 scan).  This model was then subtracted from the original visibilities using task $uvsub$.  Individual scans were then imaged for visual checking.  At this step, 2 out of 3 scans in the 2018 run were flagged out, as they had features which were apparently 180 degree symmetric, indicative of the Hermitian nature of interferometric sidelobes.  These 2 scans were conducted on the same day, and had larger phase variations compared to the other 7 scans.
After further flagging of spectral window edges and spurious features, the point source-subtracted visibilities were combined with task $concat$ and imaged together.
  Panel d) in figure \ref{image} shows the point source-subtracted image.  The image noise is $1\sigma = 0.11\ \mathrm{mJy/beam}$, with a beam size of $3^{\prime \prime}.6 \times 3^{\prime \prime}.2$.  This corresponds to $\mathrm{DR}\sim 85000$.  The theoretical noise of the data, calculated using the effective bandwidth and assuming that the noise is dominated by thermal processes, was $2\ \mathrm{\mu Jy/beam}$ over an effective bandwidth of 6.5 GHz.  Therefore, the image is still DR limited.
  An extended component is apparent around the central source at a $> 5\sigma$ level over a linear scale of $\sim 10^{\prime \prime}$, corresponding to $\sim 27\ \mathrm{kpc}$.

\subsection{Band 6} \label{section_b6}
ALMA observations at 233 GHz (Band 6) were obtained in Cycle 7 (2019.1.00634) in order to verify the existence of extended emission detected in band 3 archival data.

Band 6 observations were conducted with the ACA 7m and 12m arrays simultaneously, within a timeframe of 20 minutes.  This ensured that intrinsic flux variations of the source were negligible, and that the ACA and 12m array can be calibrated together.  After standard pipeline calibration, an initial image
showed that the peak flux at 233 GHz was 4.04 $\mathrm{Jy/beam}$ with $1\sigma = 13\ \mathrm{mJy/beam}$ ($\mathrm{DR} \sim 300$), with a beamsize of $1^{\prime \prime}.4 \times 1^{\prime \prime}.2$ (panel e) of figure \ref{image}).

Four iterations of self calibration were performed with solution intervals of 10s, 5s only for the phases, then with solution interval of 1s with both amplitude and phase, and then lastly with $solnorm=T$.  The final peak flux was 4.67 $\mathrm{Jy/beam}$, but did not change during the last 2 iterations.  The resulting noise decreased to $1\sigma = 0.2\ \mathrm{mJy/beam}$.

Subtraction of the central point source was done in the same way as band 3.  Panel f) of figure \ref{image} shows the point source subtracted image.  The image noise is $1\sigma = 0.12\ \mathrm{mJy/beam}$.  This corresponds to $\mathrm{DR}\sim 39000$.  The theoretical noise of the data was $7.9\ \mathrm{\mu Jy/beam}$ over an effective bandwidth of 8.0 GHz.  The image is still DR limited.
An extended component can be seen, elongated in the northeast-southwest direction at the $ > 2\sigma$ level.  Two compact components adjacent to the central source are significant at the $>20 \sigma$ level, but these components could arise from insufficient subtraction of the central point source.  As the point spread function (PSF) at the center with a peak flux of 4.67 $\mathrm{Jy/beam}$ decreases to the $1\sigma$ noise level at a radius of approximately $2^{\prime \prime}$, we disregard the central $4^{\prime \prime}$ from further analyses.

\subsection{Band 7}
ALMA 343 GHz (Band 7) data were obtained together with Band 6.  For Band 7, observations were conducted with the ACA only, with 9 antennas.  Observations were completed within a timeframe of 90 minutes.  
After standard pipeline calibration, an initial image
showed that the peak flux at 343 GHz was 3.13 $\mathrm{Jy/beam}$ with $1\sigma = 17\ \mathrm{mJy/beam}$ ($\mathrm{DR} \sim 180$), with a beamsize of $5^{\prime \prime}.8 \times 3^{\prime \prime}.4$ (panel g) of figure \ref{image}).

Four iterations of self calibration were performed with solution intervals of 60s, 5s only for the phases, then with solution interval of 1s with both amplitude and phase, and then lastly with $solnorm=T$.  The final peak flux was 3.21 $\mathrm{Jy/beam}$ with $1\sigma = 1.5\ \mathrm{mJy/beam}$, with a beamsize of $5^{\prime \prime}.8 \times 3^{\prime \prime}.4$.

Point source subtraction was done in the same way as band 3.  Panel h) of figure \ref{image} shows the point source subtracted image.  The image noise is $1\sigma = 1.3\ \mathrm{mJy/beam}$, with a beam size of $6^{\prime \prime}.0 \times 3^{\prime \prime}.4$.  This corresponds to $\mathrm{DR}\sim 2500$.  The theoretical noise of the data was $81\ \mathrm{\mu Jy/beam}$ over an effective bandwidth of 8.0 GHz.  The image is still DR limited.

Residual emission can be seen in the center, but no significant extended emission was found.  

\begin{figure*}[htb!]
\includegraphics[clip, height=10.5cm, trim=-1.5cm 0cm 0cm 1cm]{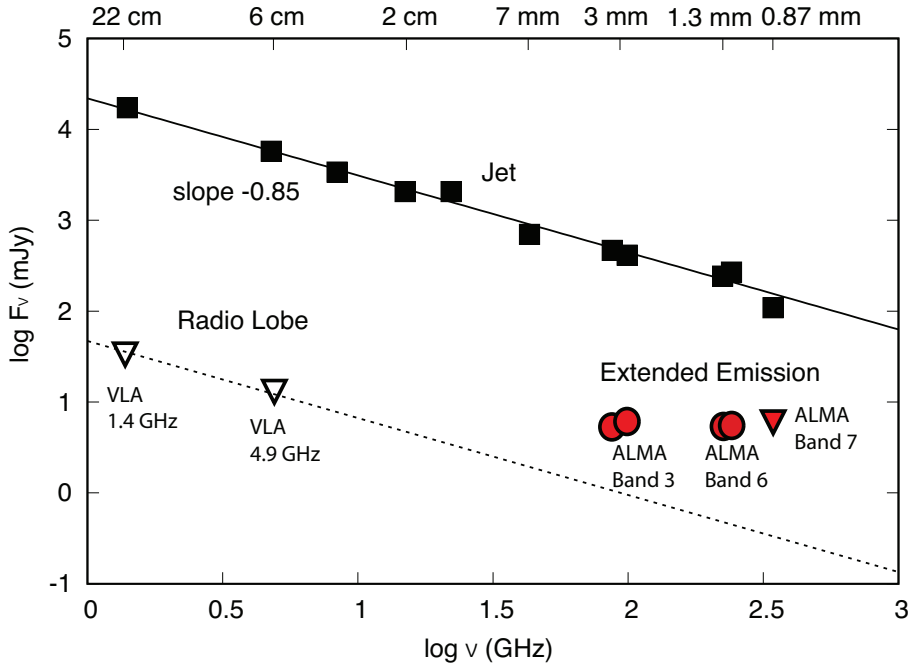}
\caption{Spectral energy distribution of 3C273.  Filled red circles are for the
extended component discussed in section \ref{extended}.  For band 7, the triangle indicates the $3\sigma$ upper limit.
Filled squares are for the jet component measured in the region specified in panel b) of figure 1.  The solid line is a power law fit to the jet component.
Open triangles are the upper limit estimates of the radio lobe component explained in section \ref{synch1}, along with the power law fit to the jet component moved parallel to match the radio lobe flux to guide the eye.}
\label{sed}
\end{figure*}

\section{Spectral energy distribution}
\subsection{Jet}\label{jet}
The most prominent structure in the ALMA bands after the subtraction of the central source, is the synchrotron jet emitted from the central source in the southwest direction.
The jet itself is orders of magnitude weaker compared to the central peak, and invisible at higher frequency ALMA bands when only the standard imaging procedure is applied, due to the low dynamic range.  We measured its total flux to compare with centimeter flux obtained with the VLA.  The region where flux was measured is shown in panel b) of figure \ref{image}.  The region is selected to include the bulk of the millimeter and centimeter jet, but the boundary of the jet in the vicinity of the nuclear source cannot be determined unambiguously.  A millimeter knot can be seen at the boundary of the jet region in panels d) and f) of figure \ref{image}, but its flux is $\sim 2\ \mathrm{mJy}$, which accounts for less than 1\% of the total jet flux, thus its effect is negligible for the discussion presented here.

Table \ref{diffuse_flux} lists the VLA and ALMA flux, and the corresponding spectral energy distribution (SED) is shown in figure \ref{sed}.
The jet SED can be fit well by a simple power law of the form $F_\nu \propto \nu^{\alpha}$, where $\alpha = -0.85\pm 0.03$.  The reduced chi square of the fit was 0.007.

\begin{table}
\caption{Flux measurements of 3C273}\label{diffuse_flux}
  \begin{center}
    \begin{tabular}{ccl} \hline
      Frequency &  Flux  & Reference \\ 
        GHz	       &  mJy  &    \\ 
        (1)		&  (2)   &    (3)       \\ \hline \hline
        	 		&  Extended &   \\
	1.4		&	$<3.4\times 10^1$		& 1\\
        4.9		&	$<1.4\times 10^1$	     	& 1\\
        87		&	$5.3 \pm 0.4$		& 2 : B3 LSB \\
        93		&	$6.0 \pm 0.3$ 		& 2 : B3 agg.\\
        99		&	$6.1 \pm 0.5$		& 2 : B3 USB \\
        225		&	$5.4 \pm 1.0$		& 2 : B6 LSB  \\
        233		&	$5.6 \pm 0.8$		& 2 : B6 agg.\\
        241		&	$5.5 \pm 1.0$		& 2 : B6 USB  \\
        343		&	$2 \pm 2$		& 2 : B7 agg.\\  \hline
           	 		&  Jet &   \\
	1.4		&	$(1.70\pm 0.01)\times 10^4$    &  1 \\
	4.9		&	$(5.80\pm 0.01)\times 10^3$    & 1\\
        8.4		&	$(3.41\pm 0.01)\times 10^3$    & 1  \\
        15		&	$(2.08\pm 0.01)\times 10^3$    & 1 \\
        22		&	$(2.05\pm 0.02)\times 10^3$    &1\\
        43		&	$(7.0\pm 0.5)\times 10^2$    & 1\\
        87		&	$(4.68\pm 0.01)\times 10^2$	& 2 : B3 LSB \\
        99		&	$(4.11\pm 0.01)\times 10^2$	& 2 : B3 USB \\
        225		&	$(2.4\pm 0.1)\times 10^2$ 	&	2 : B6 LSB  \\
        241		&	$(2.6\pm 0.2)\times 10^2$ 	&	2 : B6 USB  \\
        343		&	$(1.1\pm 0.2)\times 10^2$	&	2 : B7 agg. \\  \hline
    \end{tabular}
    \end{center}
\tablenotetext{}{Col. (1) Frequency.  ;
Col. (2) Flux integrated over an ellipse of $13^{\prime \prime}\times 6^{\prime \prime}$ for the extended emission, and integrated 
over the jet structure for the jets.  Errors are estimated by placing the same region in an emission free part of the image, and multiplying 
its $1\sigma$ noise level by the square of the number of beam elements in the region ; 
Col. (3) Reference to data used for flux measurement.  1: \citet{perley17}.  2 : This study.  B3, B6 and B7 for ALMA bands 3, 6 and 7, respectively.  
LSB and USB stand for the lower and upper sidebands of each receiver band, and ``agg." for the aggregated bands.}
\end{table}

\subsection{Extended component}\label{extended}
The high dynamic range attained by the ALMA image reveal a diffuse, extended component around the central source.  The emission is annotated in panels d) and f) of figure 1.  
The emission extends to the northeast, opposite from the jet, showing that this component is unlikely to be the root of the jet.

We measure the flux of the extended component within an ellipse with major axis $13^{\prime \prime}$, minor axis $6^{\prime \prime}$ and position angle of 0 degrees centered on the central source, which is a reasonable compromise between capturing the bulk of the extended component and avoiding the root of the jet extending to the southwest.  This region is also consistent with the area where extended [OIII] is detected in \citet{husemann19b}.
For reasons explained in section \ref{section_b6}, we subtract the flux in the central $4^{\prime \prime}$.  Table \ref{diffuse_flux} lists the measured flux of the ALMA bands.  Flux are listed separately for the lower and upper sidebands, and also for the aggregated band in bands 3 and 6.  For band 6, the flux becomes larger by approximately 20\%  if the image is smoothed to a common angular resolution as the band 3 image.
For the following analyses, we use the average flux of $5.8\ \mathrm{mJy}$ over bands 3 and 6.  For band 7, error in the flux is large and we do not detect any significant diffuse emission at this frequency.  

\section{Origin of extended emission}
A number of emission mechanisms can be significant at the ALMA bands.  In particular, in band 3, the continuum emission can generally be
a mixture of synchrotron, thermal bremsstrahlung, and the Rayleigh-Jeans tail of the dust spectrum.  The flat spectrum observed in the millimeter range
indicates that dust cannot be a responsible for bulk of the extended emission, as we would expect an increasing flux at higher frequencies for thermal dust emission.

\subsection{Synchrotron associated with nuclear activity} \label{synch1}
Extended centimeter emission, or radio lobe, has been detected in the vicinity of the central source with the VLA \citep{punsly16, perley17}.  Here we coin the term ``radio lobe'' specifically for the centimeter detection, to make the distinction from the extended millimeter component.  \citet{punsly16} discussed the radio lobe in detail, and conclude that the lobes are relaxed synchrotron cooling plasma associated with the jet and the counter jet.
In order to estimate the extended centimeter flux of the region where we measure the extended millimeter flux, we assume that the spatial extrapolation of the radio lobe flux to the central region gives its upper limit.  This is a reasonable assumption, as radio lobes are bipolar structures which peak at the lobes, created by its interaction with the intergalactic medium.  We measured the average flux density in the VLA 1.4 GHz (L band) and 4.9 GHz (C band) images for regions where the radio lobe was strongest, in the northeast direction from the nucleus (counter-jet side; region shown in panel b) of figure \ref{image}).  Then, we scaled the flux assuming that the same flux density applies for the central $13^{\prime \prime} \times 6^{\prime \prime}$.  The flux limit estimates for the VLA bands are 
listed in table 1.

From the SED of the extended millimeter component shown in figure \ref{sed}, it is clear that in the millimeter range, the SED is flat or can allow for only a very shallow ($\alpha=-0.03\pm 0.07$) spectral index.    The centimeter flux is significantly stronger than in the millimeter.  We spectrally extrapolate the radio lobe flux to the millimeter assuming $\alpha=-0.85$, the same value as the jet.  This is justified because the the nuclear source is the common origin for the relativistic electrons which give rise to the radio lobe and the jet, and the electron number distribution is likely to be similar as well.  The assumed value is also in line with the synchrotron spectral index of the KINGFISHER sample of nearby galaxies ($-0.97\pm 0.16$, \citet{tabatabaei17}).  The expected flux at 93 GHz is $\le 1\ \mathrm{mJy}$, and $\le 0.5\ \mathrm{mJy}$ at 233 GHz.  Thus, synchrotron emission originating from the nuclear source or the jet, accounts for less than 17\% of the extended 93 GHz flux, and less than 9\% for the extended 233 GHz flux, indicating that it is not the dominant mechanism responsible for extended emission in the millimeter range.

\subsection{Synchrotron associated with star formation} \label{sync2}
Synchrotron emission from galaxies can originate from cosmic rays accelerated in supernova remnants, i.e., star formation within the disk of the host galaxy.  If the extended millimeter emission is synchrotron emission tracing massive star formation, we can convert the millimeter flux to the star formation rate (SFR).  Assuming a synchrotron spectral index
of $-0.85$, the observed millimeter flux of $5.8\ \mathrm{mJy}$ corresponds to a SFR of
$\sim 3500\ \mathrm{M_\odot yr^{-1}}$ \citep{condon92}.  This is an order of magnitude higher than estimates in other radio loud quasar hosts \citep{zakamska16a, hardcastle19}.  The stellar mass of the host elliptical galaxy of 3C273 is estimated to be $M_*\sim 10^{11.1}\ \mathrm{M_\odot}$ with
an average stellar population of 2.5 Gyr \citep{zhang19}.  For nearby star forming galaxies on the main sequence, the expected SFR at this stellar mass is only a few $\times 10\ \mathrm{M_\odot yr^{-1}}$ \citep{peng10}.  Thus, if the observed millimeter flux is due to synchrotron emission from massive star formation, the host of 3C273 must be an extreme starburst, but such a bursty phase contradicts with the relatively old stellar population.
In addition, the molecular gas mass in 3C273 estimated from $\mathrm{^{12}CO}(J=1-0)$ emission is only $\sim 10^9\ \mathrm{M_\odot}$ \citep{husemann19b}, which is comparable to  local spirals which have SFR of only few $\mathrm{M_\odot yr^{-1}}$.  We therefore conclude that synchrotron originating from massive star formation in the host galaxy is not the source of extended millimeter emission in 3C273.

\subsection{Thermal Bremsstrahlung}
An alternative mechanism of radio emission is thermal bremsstrahlung.  Thermal bremsstrahlung is weak and normally swamped by synchrotron at the centimeter and by dust at the submillimeter wavelengths.  Thermal bremsstrahlung normally has a shallow spectral slope of $\alpha \sim -0.1$, which is consistent with our nearly constant flux observed in the ALMA bands.  Assuming thermal bremsstrahlung, we can convert the observed millimeter flux of $5.8\ \mathrm{mJy}$ to the production rate of ionizing photons using the approximation by \citet{rubin68}, 
\begin{equation}
\begin{split}
\left[ \frac{Q(H^0)}{s^{-1}}\right] \geq  6.3 \times 10^{25}\times \\
\left( \frac{T_e}{10^4\ \mathrm{K}}\right)^{-0.45} \left( \frac{\nu}{\mathrm{GHz}}\right)^{0.1}  \left( \frac{L_\nu}{\mathrm{erg\ s^{-1} Hz^{-1}}}\right)
\end{split}
\end{equation}
which assumes equilibrium between ionization and recombination, and the equality holds only if the system is ionization bound.  For an electron temperature of $T_e \sim 10^4\ \mathrm{K}$, we have $Q(H^0) > 4.0\times 10^{56}\ \mathrm{s^{-1}}$.  Here we have omitted the equality, since the radio jet protruding from the host galaxy indicates that the system is at least partially density bound (i.e., some ionizing photons escape from the host galaxy).  We denote this lower limit value as $Q(H^0)_\mathrm{min} = 4.0\times 10^{56}\ \mathrm{s^{-1}}$, which indicates the actual number of ionizing photons that were used to create the nebula.
There are two candidate sources of the ionizing photons, massive star formation in the host galaxy, or direct ionization of host ISM by the central source.

If the extended emission is due to massive star formation, the ionizing photon rate above corresponds to
$\mathrm{SFR} > 3000\ \mathrm{M_\odot yr^{-1}}$ \citep{condon92, murphy11, bendo16, michiyama20}.  Following the discussion presented in subsection \ref{sync2}, this SFR is unreasonably high, making thermal 
bremsstrahlung from massive star formation an unlikely origin of the extended millimeter emission.  This is consistent with results by \citet{husemann19b} that optical emission 
line ratios of the host galaxy are inconsistent with star formation.

Alternatively, the central source can directly ionize atomic gas in the host galaxy, giving rise to the thermal bremsstrahlung.  Discussion in the previous sections likely rule out other origins, leaving this ``QSO feedback'' scenario as the plausible explanation of the extended millimeter flux discovered in 3C273.  In this case, the thermal emission can be regarded as the radio counterpart of EELR observed through optical forbidden lines.

\citet{martel03} observed 3C273 with the HST coronagraph, and find that the outer halo of the host galaxy is lopsided to the northeast direction, out to approximately $10^{\prime \prime}$ ($1^{\prime \prime} = 2.7\ \mathrm{kpc}$).  The EELR detected in [OIII] by \citet{husemann19b} find that the  [OIII] emission extends typically to $7^{\prime \prime}$.
Our detection of the extended millimeter component extends to radii of $\sim 7^{\prime \prime}$, similarly to the EELR, while extending to the northeast direction.  The distribution of ionized gas detected in thermal bremsstrahlung is thus consistent with that detected in [OIII].

\section{negative feedback in the host galaxy of 3C273}

The luminosity used to ionize atomic hydrogen, $E_\mathrm{feedback}$, can be estimated by 
\begin{equation}
E_\mathrm{feedback} > Q(H^0)_\mathrm{min} \varepsilon_\mathrm{Ly} = 8.8\times 10^{45}\ \mathrm{erg\ s^{-1}}
\end{equation}
 where $\varepsilon_\mathrm{Ly}=2.2\times 10^{-11}\ \mathrm{erg}$ is the ionization energy of hydrogen.  Here the value is given as a lower limit because the incident ionizing photons could have higher energy than $\varepsilon_\mathrm{Ly}$, in which case the residual energy after ionization is reflected on the kinetic energy of the electrons, therefore included as negative feedback energy into the host ISM.
 The total ionizing flux of the central source is estimated to be $3.9\times 10^{46}\ \mathrm{erg\ s^{-1}}$ \citep{malkan82}, and the bolometric flux to be 
 $1.3\times 10^{47}\ \mathrm{erg\ s^{-1}}$ \citep{shang05}.
Assuming that 3C273 has not changed significantly in brightness over the past $\sim 10^5\ \mathrm{yr}$ (i.e, the extent of the extended emission in light years), this implies that approximately $> 7\%$ of the bolometric flux, and $> 20\%$ of the total ionizing flux from 3C273, is fed back into the ISM of the host galaxy via ionization and electron heating.

We can also estimate the mass of ionized gas, $M_\mathrm{HII}$ by 
\begin{equation}
M_\mathrm{HII} = Q(H^0)_\mathrm{min} \mathrm{m_H} \tau_\mathrm{H}
\end{equation} 
where $\mathrm{m_H}$ is the hydrogen mass, and $\tau_\mathrm{H}$ is the recombination time of hydrogen.  $\tau_\mathrm{H}$ is given by 
\begin{equation}
\tau_\mathrm{H}=\left(n_e R_\mathrm{H}\right)^{-1}
\end{equation}
where $n_e$ is the electron density and  $R_\mathrm{H}=3\times 10^{-13}\ \mathrm{cm^3\ s^{-1}}$ is the hydrogen recombination rate for Case B recombination \citep{osterbrock89}.
If we assume the electron density to be $n_e = 10^{1-2}\ \mathrm{cm^{-3}}$ \citep{steidel14, kaasinen17, lee19}, applicable to local star forming galaxies, we obtain a hydrogen recombination time of $10^{3-4}\ \mathrm{yr}$ and $M_\mathrm{HII} = 10^{10-11}\ \mathrm{M_\odot}$.  This places the ionized gas to stellar mass ratio in the range $\frac{M_\mathrm{HII}}{M_*} \sim 0.1 - 1$, higher than that in the Milky Way which has $\frac{M_\mathrm{HII}}{M_*} \sim 0.04$ \citep{walterbos98, nakanishi16, licquia16}.  It is unclear whether the massive ionized gas estimated here indicates negative feedback at work, however.  \citet{osullivan15} find that elliptical galaxies with $M_* \sim 10^{11}\ \mathrm{M_\odot}$ have molecular gas of $M_\mathrm{H2} \sim 10^{8-9}\ \mathrm{M_\odot}$, which is consistent with the estimate in 3C273 \citep{husemann19b}.  Thus the host of 3C273 is not particularly molecular gas deficient, at least with respect to other massive ellipticals which also tend to harbour AGN.  
Either the central QSO has ionized a respectable fraction of neutral gas in this galaxy but not enough to suppress molecular gas formation, or molecular gas formation is actually inhibited due to ionization of available neutral gas, but we are witnessing the short time after molecular gas formation has stopped but has not yet been consumed by star formation.  The latter scenario is viable.  The SFR in ellipticals with similar stellar mass range are typically smaller than $10\ \mathrm{M_\odot yr^{-1}}$ \citep{osullivan15}, so the molecular gas consumption timescale in 3C273 is likely longer than $\sim 10^8\ \mathrm{yr}$.
The age of the synchrotron jet in 3C273 is estimated to be $\sim 10^5\ \mathrm{yr}$ \citep{stawarz04}, consistent with the size of the extended thermal bremsstrahlung emitting region ($10^5$ light years), so 3C273 may be at the very start of its QSO phase.  Although 3C273 may be strong enough to ionize atomic gas throughout the host, it may not have lasted long enough to have had any apparent negative feedback impact on the host galaxy yet.  If, however, the current activity of 3C273 continues for a timescale comparable to $\sim 10^8\ \mathrm{yrs}$ \citep{yu02}, star formation will die down due to lack of molecular gas, and we will observe that star formation has been suppressed due to negative feedback.  Further high dynamic range observations of the atomic gas phase, and estimation of the SFR in the host galaxy, will be able to shed more light on the evolution of 3C273.

\section{Summary}
We have used archival and newly obtained ALMA data in bands 3, 6 and 7 to image 3C273, an archetypical radio loud QSO at $z=0.158$.
After self calibration and point source subtraction, we reach a dynamic range of 85000 at band 3, 39000 at band 6 and 2500 at band 7. 
Although the images are dynamic range limited, it demonstrates the capability for high image dynamic range using ALMA.  We detect extended millimeter continuum emission associated with the host galaxy, that is spatially consistent with
EELR detected in [OIII].  The flux of the extended emission and its comparison with VLA data show that the extended emission cannot be explained 
by synchrotron emission, or thermal bremsstrahlung due to massive star formation within the ISM of the host galaxy.  
We interpret the extended emission in the context of thermal bremsstrahlung from gas directly ionized by the central source.
Energetics suggest that more than 7\% of the bolometric flux, and more than 20\% of the total ionizing flux of 3C273, is used to ionize neutral hydrogen in the host.
A rough estimate of the ionized gas mass indicates that 3C273 has ionized a significant amount of gas, but either not enough to suppress molecular gas formation, or we are witnessing a short period in the lifetime of a QSO-host interaction where molecular gas formation has stopped due to ionization of atomic gas, but
star formation has not consumed molecular gas yet, so that the negative feedback effect is not yet observable.
Future work should include estimates of the neutral gas mass and SFR in this prototypical QSO.  We find that
observing the radio counterpart of EELRs circumvents the difficulties related to assumptions in dust extinction and emission line excitation mechanisms associated to observation of optical emission lines, and thereby presents a valuable alternative to studying QSO feedback in host galaxies.

\begin{acknowledgments}
The authors thank the anonymous referee for helpful comments which improved the manuscript greatly.
This paper makes use of the following ALMA data: ADS/JAO.ALMA\#2017.1.00815.S and ADS/JAO.ALMA\#2019.1.00634.S. ALMA is a partnership of ESO (representing its member states), NSF (USA) and NINS (Japan), together with NRC (Canada), MOST and ASIAA (Taiwan), and KASI (Republic of Korea), in cooperation with the Republic of Chile. The Joint ALMA Observatory is operated by ESO, AUI/NRAO and NAOJ.  This research is supported by JSPS KAKENHI Grant Number JP 20K04015 and 19K14759.
\end{acknowledgments}

%

\vspace{5mm}
\facilities{ALMA}


\software{CASA \citep{casa}}



\bibliography{sample631}{}
\bibliographystyle{aasjournal}



\end{document}